\documentclass{llncs} \def\versy{cie}

\def\cie{cie}
\def\fully{full}
\ifx\versy\cie

\fi
\ifx\versy\fully

\fi

\usepackage{amsmath}
\usepackage{amsfonts}
\usepackage{amssymb}
\usepackage{amscd}
\usepackage{verbatim}
\usepackage{url}

\spnewtheorem{thm}{Theorem}{\bf }{\it }

\spnewtheorem{defin}[thm]{Definition}{\bf}{\rm }

\spnewtheorem{prop}[thm]{Proposition}{\bf}{\it}

\spnewtheorem{open}[thm]{Open question}{\bf}{\it}

\spnewtheorem{lem}[thm]{Lemma}{\bf}{\it}

\spnewtheorem{cor}[thm]{Corollary}{\bf}{\it}

\spnewtheorem*{rem}{Remark}{\it}{}

\numberwithin{equation}{section}

\newcommand{\N}{\mathbb{N}}
\newcommand{\cs}{\{0,1\}^\omega}
\newcommand{\fs}{\{0,1\}^*}
\newcommand{\emptystring}{\epsilon}
\newcommand{\jump}{\mathbf{0}'}

\newcommand{\R}{\mathbb{R}}

\newcommand{\ivm}{\mathfrak{D}}

\newcommand{\permit}{\ensuremath{\texttt{permit}}}
\newcommand{\uh}{\mathrel{\upharpoonright\nolinebreak[4]
  \hspace{-0.65ex}\upharpoonright}}

\renewcommand{\d}{\mathbf{d}}
\newcommand{\converge}{\mathop\downarrow}
\renewcommand{\complement}[1]{\overline{#1}}

\newcommand{\emphdef}[1]{\emph{#1}}

\newcommand{\size}[1]{{\left|#1\right|}}

\title{How powerful are integer-valued martingales?}

\author{Laurent Bienvenu\inst{1},
Frank Stephan\inst{2} and
Jason Teutsch\inst{3}}
\institute{LIAFA, CNRS \& Universit\'e de Paris 7, France {\tt
laurent.bienvenu@liafa.jussieu.fr}\\
\and
National University of Singapore \\{\tt fstephan@comp.nus.edu.sg}
\and Center for Communications Research--La Jolla, USA \\{\tt
jrteuts@ccrwest.org}
}

\begin{document}

\maketitle

\begin{abstract}
In the theory of algorithmic randomness, one of the central notions is
that of computable randomness. An infinite binary sequence~$X$ is
computably random if no recursive martingale (strategy) can win an
infinite amount of money by betting on the values of the bits of~$X$.
In the classical model, the martingales considered are real-valued,
that is, the bets made by the martingale can be arbitrary real
numbers. In this paper, we investigate a more restricted model, where
only integer-valued martingales are considered, and we study the class
of random sequences induced by this model.
\end{abstract}

\section{Gambling with or without coins}

One of the main approaches to define the notion of random sequence is the so-called ``unpredictability paradigm''. We say that an infinite binary sequence is ``random'' if there is no effective way to win arbitrarily large amounts of money by betting on the values of its bits. The main notion arising from this paradigm is computable randomness, but other central notions such as Martin-L\"of randomness, Schnorr randomness, and Kurtz randomness, can be formulated in this setting. For all of these notions, we consider models of games where the player can, at each turn, bet \emph{any} amount of money between~$0$ and his current capital. In ``practice'' however, one cannot go into a casino and bet arbitrarily small sums of money: there is always a unit value, and any bet made has to be a multiple of this value. Some casinos (and games) also impose upper limits on the amount of capital the one can gamble in each round of play.  In the following exposition, we examine the consequences of restricting betting amounts to integers and finite sets.\\

%

To formalize the unpredictability paradigm, we need the central notion of \emph{martingale}. A martingale is a betting strategy for a fair game and is formally represented by a function that corresponds to the gambler's fortune at each moment in time. Let $\fs$ denote the set of
all finite binary sequences, and $\cs$ is the set of all countably
infinite binary sequences (a.k.a\ \emphdef{reals}).  Any function $M : \fs \to \R^+$ which
satisfies the \emphdef{fairness condition}
\begin{equation} \label{eq:fairness condition}
M(\sigma) = \frac{M(\sigma0) + M(\sigma1)}{2}
\end{equation}
for all $\sigma \in \fs$ is called a \emphdef{martingale}. $M(\sigma)$ corresponds to the gambler's capital after having already bet on the finite sequence $\sigma$.  The fairness condition~\eqref{eq:fairness condition} says that the amount of
money gained from an outcome of ``0'' is the same that would be lost
from an outcome of ``1''.  It is important to note that our definition of martingale is a very restricted version of what is usually referred to as ``martingale'' in probability theory, where it is defined to be a sequence $X_0, X_1, \ldots$ of real-valued random variables (possibly taking negative values) such that for all~$n$
\[
\mathbb{E}[X_{n+1} | X_0, X_1, \ldots, X_n] = X_n.
\]
To make the distinction, we call such a sequence a \emphdef{martingale process}. A martingale is called \emphdef{recursive} if $M$ is a recursive function.  Throughout this exposition, ``martingale'' and ``recursive martingale'' will be used
synonymously.

For any $A \in \cs$, $A \uh n$ is the finite binary sequence, or
\emphdef{initial segment}, consisting of the first~$n$ digits of $A$. $\complement{A}$ denotes the complement of a set $A$ (when seen as a subset of $\N$).
We also identify sets with their characteristic sequences. $\size{\sigma}$ denotes the length of a binary sequence $\sigma$. A martingale $M$ \emphdef{succeeds} on $A \in \cs$ if $M$ achieves arbitrary sums of money over $A$, that is, $\lim\sup_n M(A \uh n) = \infty$. Otherwise $A$ \emphdef{defeats} $M$. $M$ \emphdef{Schnorr-succeeds} on a set $A$ if $M$ succeeds on $A$ and there exists a recursive, non-decreasing, unbounded function~$f$ such that $f(n) < M(A \uh n)$ for infinitely many $n$. $M$ \emphdef{Kurtz-succeeds} on a set~$A$ if $M$ succeeds on $A$ and there exists a recursive, non-decreasing, unbounded function~$f$ such that $f(n) < M(A \uh n)$ for all~$n$.  We can now define the main classical notions of randomness in terms of martingales.

\begin{defin}
 A sequence $A \in \cs$ is called \emphdef{computably random} if
$A$ defeats every martingale.
If no martingale Schnorr-succeeds on $A$, then $A$ is \emphdef{Schnorr
random}.  If no martingale Kurtz-succeeds on $A$, then $A$ is \emphdef{Kurtz
random} (equivalently, $A$ is Kurtz random if and only if $A$ does not belong to any $\Pi^0_1$ subset of $\cs$ of measure 0).
\end{defin}


%

In this paper, we shall consider games where the player can only make bets of integer value. For $M$ a martingale and $\sigma \in \fs$,
$\size{M(\sigma0) - M(\sigma)}$ is called the \emphdef{wager at $\sigma$}. Now, given a set~$V$ of non-negative integers, we say that a martingale is \emphdef{$V$-valued} if for all $\sigma$ the wager of $M$ at $\sigma$ belongs to $V$, unless $M$ does not have enough capital in which case the wager at $\sigma$ is~$0$. Formally, $M$ is $V$-valued if for all $\sigma \in \fs$ and $a \in \{0,1\}$, $M(\sigma) < \min(V) \Rightarrow M(\sigma a)=M(\sigma)$ and $M(\sigma) \geq \min(V) \Rightarrow |M(\sigma a)-M(\sigma)| \in V$. A martingale whose wagers are integers is called an \emph{integer-valued} martingale. In case $V$ is finite we say that $M$ is \emphdef{finitely-valued} and if~$V$ is a singleton, that $M$ is \emphdef{single-valued}.

\begin{defin}
A real $X$ is \emphdef{$V$-valued random} if no $V$-valued
martingale succeeds on $X$.  A real $X$ is a
\emphdef{finitely-valued / integer-valued / single-valued random} if
no finitely-valued / integer-valued / single-valued martingale
succeeds on $X$.
\end{defin}

The rest of the paper studies how these new notions of randomness interact with the classical ones. We will prove the implications of the following diagram:\\

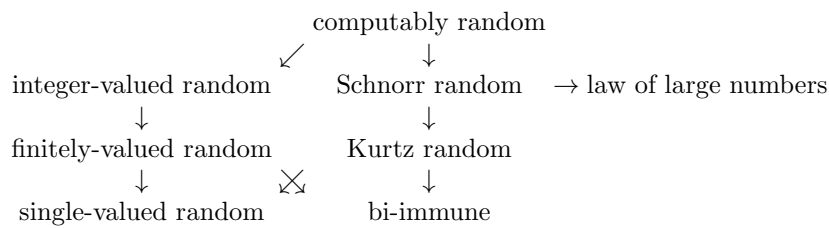
\begin{center}
\begin{tabular}{ccccc}
 & &computably random & & \\
 & $\swarrow$ & $\downarrow$ \\
integer-valued random & & Schnorr random & $\rightarrow$ &  law of
large numbers\\
 $\downarrow$ & & $\downarrow$ \\
 finitely-valued random &  & Kurtz random & \\
 $\downarrow$ & $\searrow$ \hspace{-4ex} $\swarrow$ & $\downarrow$ & \\
 single-valued random & & bi-immune & & \\
\end{tabular}
\end{center}
and we shall further see that no other implication than those indicated (and their transitive closure) holds. We than an infinite set is called \emphdef{immune} if it contains no infinite r.e.\ set.  Even stronger, an infinite set $A = \{a_0 < a_1 < a_2 <
\dotsb \}$ is \emphdef{hyperimmune} if there exists no recursive function $f$ such that $f(n) > a_n$ for all $n$ \cite{Soare1987}.  A
(hyper)immune set whose complement is also (hyper)immune is called \emphdef{bi-(hyper)immune}.  Thus a member of $\cs$ is bi-immune if and only if there no recursive list of positions containing all 0's or all 1's.  A set $A$ is $\Sigma_0^n$ if $A$ can be defined using a formula with $n$ quantifiers followed by a recursive predicate where the leading quantifier is existential.  $\jump$ denotes the halting set, and we say \emphdef{$A$ is $B$-recursive} if $A$ is Turing reducible to $B$.

 For additional background on algorithmic
randomness, see the forthcoming book of Downey and Hirschfeldt
\cite{DH2010} and the new book of Nies \cite{Nies2009}.\\

If we were to ask someone what the absolute
minimum one could expect from a set called ``random,'' you might
receive one of the following two responses:
\begin{quotation}
1. The set obeys the law of large numbers.

2. The set is bi-immune.
\end{quotation}
The person who says ``1'' believes that a set which does not follow
the law of large numbers exhibits a probabilistic bias in its
distribution of 0's and 1's.  The person who says ``2'' believes that
a set with an infinite recursive subset of 0's or 1's yields
algorithmic bias.  There exists, however, a third possibility:\begin{quotation}
3. The set is single-valued random.
\end{quotation}
``3'' closely matches our intuition in the sense that one should not
be able to predict successive outcomes resulting from a ``random''
process.  From a practical point-of-view, single-valued randomness
also makes sense.  If you have to sit out $2^{1000}$ rounds of
roulette before placing a sure bet, as might occur when gambling on a
non-bi-immune set, then with probability 1 the casino has already
closed while you were waiting for this opportunity.  In
Section~\ref{sec:fvm}, we shall prove that notion ``3'' indeed differs
from notions ``1'' and  ``2.''\\

The separation of Kurtz randomness and Schnorr randomness is folklore
(we will see in a moment how it can be proven).  A somewhat more
difficult result is the separation of computable randomness and
Schnorr randomness. The separation of these two notions was proven by
Wang who constructed a Schnorr random sequence~$X$ together with a
martingale~$M$ that succeeds on~$X$. It turns out that in Wang's
construction, the martingale~$M$ is already $\{0,1\}$-valued, hence it
immediately follows that Schnorr randomness (a fortiori Kurtz
randomness) does not imply finitely-valued randomness (and a fortiori integer-valued randomness).
\begin{thm}[Wang~\cite{Wang1999}]
There exists a Schnorr random $X \in \cs$ and a $\{0,1\}$-valued
martingale $M$ such that $M$ succeeds on~$X$.
\end{thm}
In Section~\ref{sec:genericity} we shall see that conversely,
integer-valued randomness does not imply Schnorr randomness, and a fortiori computable randomness.

\section{Integer-valued martingales and genericity} \label{sec:genericity}

There is an essential difference between rational-valued and
integer-valued martingales. The latter can always be permanently
defeated while in general the former cannot be. Consider the example
of a player starting with an initial capital of~$1$ who at each turn
bets half of its capital on the value~$1$ (that is, the corresponding
martingale~$M$ satisfies $M(\sigma0)=M(\sigma)/2$ and
$M(\sigma1)=3M(\sigma)/2$ for all~$\sigma \in \fs$). This is a
rational-valued martingale with the following property. Pick a
stage~$s$ of the game; no matter how unlucky the player has been
before that stage, she always has a chance to recover. More precisely,
for any finite sequence of outcomes~$\sigma \in \fs$, no matter how
small $M(\sigma)$ is, the player can still win the game if the
remaining of the outcomes contains a lot of $0$'s (for example the
player wins against the sequence $\sigma0000 \ldots$). This phenomenon
no longer holds for integer-valued martingales, and in fact the
opposite is true, that is,  no matter how lucky the player has been up
to stage~$s$, there is always a risk for her to see her strategy
permanently defeated at some stage $s'>s$. This is expressed by the
following lemma.

\begin{lem}\label{lem:comeagerness}
Let $M$~be an integer-valued martingale. For any $\sigma \in \fs$,
there exists an extension $\tau(\sigma,M) \in \fs$ of $\sigma$ such
that $M(\tau')=M(\tau(\sigma,M))$ for all extensions $\tau'$ of
$\tau(\sigma,M)$ (in particular the strategy~$M$ does not succeed on
any $X \in \cs$ extending~$\tau(\sigma,M)$).
\end{lem}

\begin{proof}
Let $M, \sigma$ be fixed. We construct the string
$\tau=\tau(\sigma,M)$ via the algorithm:\\
$ $\\
1. Set $\tau \leftarrow \sigma$\\
2. While there exists an extension $\tau'$ of $\tau$ such that
$M(\tau') < M(\tau)$\\
\phantom{XXXX} Choose any such $\tau'$ and set $\tau \leftarrow \tau'$
(and go back to step 2.)\\
3. Return($\tau$)\\
$ $\\
Note that this is algorithm in a general sense, that is, we do not
claim that it can be implemented in a computable way (and indeed it
cannot be, because the condition of the ``While'' loop needs to check
the values $M(\tau')$ for all extensions of $\tau$ and there are
infinitely many of them), but only that it outputs a correct value of
$\tau$. First, to see that the algorithm terminates, notice that after
each execution of the While loop, the value of $M(\tau)$ is decreased,
and because~$M$ has integer values, this means that $M(\tau)$ is
decreased by at least~$1$. Therefore the While loop is executed at
most~$k=M(\sigma)$ times. We also claim that the output $\tau$ is
correct: indeed it must fail the condition of the While loop, that is,
for all extensions $\tau'$ of $\tau$ one has $M(\tau') \geq M(\tau)$.
But the fairness condition of martingales implies that in that case,
$M(\tau')=M(\tau)$ for all extensions~$\tau'$ of $\tau$ (this can be
checked by a straightforward induction). \qed
\end{proof}

\noindent
From a topological perspective, the above result shows that any
integer-valued martingale~$M$ is defeated on a dense open set. Indeed,
for any $\sigma$,~$M$ is defeated by every sequence~$X \in
[\tau(\sigma,M)]$ hence~$M$ is defeated by any sequence in the dense open set
\[
\mathcal{U}_M = \bigcup_{\sigma \in \fs} [\tau(\sigma,M)]
\]
(it is dense as for any $\sigma$, $[\tau(\sigma,M)] \subseteq
[\sigma]$ by construction). Therefore, the set of integer-valued
random sequences contains the intersection over all integer-valued
martingales $\bigcap \mathcal{U}_M$.  This is a countable intersection
of dense open sets, hence the following corollary.
\begin{cor} \label{cor:comeagerness2}
The set of integer-valued random sequences is co-meager.
\end{cor}
This shows that as a notion of randomness, integer-valued randomness
is quite weak. Indeed, one of the most basic properties that we
can expect from a random sequence~$X$ is that it satisfies the \emphdef{law of
large numbers}, that is, the number of $0$'s in $X \uh n$ is
$n/2+o(n)$. It is a routine exercise to show that the set of
sequences~$X$ satisfying the law of large numbers is a meager set
(contained in a countable union of closed set with empty interior).
Therefore, in the sense of Baire category, most sequences are
integer-valued random but do not satisfy the law of large numbers. On
the other hand, it is well-known that any Schnorr random sequence must
satisfy the law of large numbers \cite{Nies2009}, which yields a further corollary.

\begin{cor} \label{cor:dm-not-schnorr}
There exists a sequence~$X \in \cs$ which is integer-valued random but
not Schnorr random.
\end{cor}

If we now want to compare integer-valued randomness and Kurtz
randomness, the above results are insufficient, as the set of Kurtz
random sequences is also a co-meager set. We will prove that Kurtz
randomness does not imply integer-valued randomness by looking at the
classical counterpart of Baire category, namely genericity. Recall
that a set $W \subseteq \fs$ is \emph{dense} if the open set $\bigcup_{\sigma
\in W} [\sigma]$ is dense or equivalently if for any string $\sigma$
there exists a string in $W$ extending~$\sigma$. We say that $X \in \cs$
is \emph{weakly $n$-generic} if $X$ has a prefix in every dense $\Sigma^0_n$ set. We further say that~$X$ is
\emph{$n$-generic} if for any (not necessarily
dense) $\Sigma^0_n$ set of strings~$W$, either~$X$ has a prefix in~$W$ or there exists a
prefix of~$X$ which has no extension in~$W$. For all~$n \geq
0$ it holds that
\[
\text{weakly (n+1)-generic ~ $\Rightarrow$~ n-generic ~$\Rightarrow$ ~
weakly n-generic}.
\]
Kurtz showed that weakly 1-genericity is enough to ensure Kurtz randomness.
\begin{prop}[Kurtz~\cite{Kurtz1981}]
Any weakly 1-generic sequence~$X \in \cs$ is Kurtz random.
\end{prop}

\noindent
The next two theorems show that more genericity is needed to ensure
integer-valued randomness.  That is, weak 2-genericity is sufficient,
but 1-genericity is not.
\begin{thm}
Let $X \in \cs$ be any weakly 2-generic sequence. Then $X$ is
integer-valued random.
\end{thm}

\begin{proof}
We have shown in Lemma~\ref{lem:comeagerness} that for any martingale
$M \in \ivm$, the set of strings
\[
W_M=\{\sigma \; : \; \text{$M(\sigma')=M(\sigma)$ for all
extensions~$\sigma'$ of $\sigma$}\}
\]
is dense. It is also easy to see that this set is recursive in $\jump$, in particular $W_M$ is $\Sigma^0_2$. By definition, a
weak-2-generic sequence $X$ must have a prefix in $W_M$ for all
integer-valued martingales~$M$, and it is clear that if $X$ has a
prefix in $W_M$, $M$ does not succeed on~$X$. \qed
\end{proof}

\begin{thm} \label{thm:1-generics-not-dm}
There exists a $1$-generic sequence $X \in \cs$ and a $\{0,1\}$-valued
martingale $M$ such that $M$ succeeds on~$X$.
\end{thm}

\begin{proof}
We will build the sequence $X$ by constructing an increasing (for the
prefix order) sequence $(\gamma_n)$ of strings, then taking $X$ to be
the unique element of $\cs$ having all of the $\gamma_n$ as prefixes.  The martingale we construct will be $\{0,2\}$-valued, however a successful $\{0,1\}$-valued martingale can also be achieved by cutting the $\{0,2\}$-valued wagers in half. 

Let $(W_e)_{e \in \N}$ be an effective enumeration of all $\Sigma^0_1$
sets of strings. For all~$e$, set
\[
F_e= \{\tau\; : \; \exists \sigma \in W_{e,|\tau|}~ \text{and $\tau$
extends $\sigma$}\}
\]
and $F^{\min}_e$ the set of minimal elements of $F_e$, that is, the
set of $\tau$ such that $\tau \in F_e$ and no strict prefix of $\tau$
is in $F_e$. Note that whenever a string $\sigma$ is in $F_e$, then so
are all strings that extend $\sigma$, and whenever a string $\sigma$
is $F^{\min}_e$, then no strict extension of $\sigma$ is.
It is clear that the $F_e$ and $F^{\min}_e$ are (uniformly) recursive
sets, and also easy to see that a sequence $Y \in \cs$ is 1-generic if
and only if for all~$e$, either~$Y$ has a prefix in $F_e$ (resp.
$F^{\min}_e$) or some prefix of~$Y$ has no extension in $F_e$. \\

We start by defining the martingale~$M$ which will succeed on the
sequence~$X$. It is defined by $M(\emptystring)=12$ and for all
$\sigma \in \fs$ and $a \in \{0,1\}$:
\[
M(\sigma a)=\left\{
\begin{tabular}{ll}
$M(\sigma)+2$~ & \text{if $M(\sigma) \geq \permit(\sigma)$ and $a=1$}\\
$M(\sigma)-2$~ & \text{if $M(\sigma) \geq \permit(\sigma)$ and $a=0$}\\
$M(\sigma)$~ & \text{if $M(\sigma) < \permit(\sigma)$}
\end{tabular}
\right.
\]

where the function $\permit$ is defined inductively by
\begin{align*}
\permit(\emptystring)&=4; \\
\permit(\sigma a) &= \min \left[\{\permit(\sigma)+1\}\cup\{4e+4 : \sigma a \in F^{\min}_e\}\right].
\end{align*}
for $\sigma \in \fs$ and $a \in \{0,1\}$.
It is also easy to see that $\permit$ and~$M$ are recursive, and $M$ is
integer-valued (its values are positive because $\permit(\sigma) \geq
4$ for all~$\sigma$, hence $M$ is never allowed to make a bet if its
capital is less than~$4$).

We now define the sequence of strings $(\gamma_n)$, and an auxiliary
sequence $(\zeta_n)$ by setting $\gamma_0=\emptystring$,
$\zeta_0=\emptystring$ and inductively, for all~$n$:
\begin{itemize}
\item[(a)] if there exists an extension of $\gamma_n$ in $F_n$ then
let $\zeta_{n+1}$ be a shortest such extension (chosen effectively), and
\item[(b)] if there exists no such extension, let $\zeta_{n+1}=\gamma_n$.
\end{itemize}
Finally, define $\gamma_{n+1}=\zeta_{n+1} 11111111$.\\

Note that $\zeta_n$ can be determined from $\gamma_n$ using the oracle
$\jump$, hence the sequence $X$ obtained in this construction (by
taking the limit of the $\gamma_n$, or equivalently the limit of the
$\zeta_n$) is also recursive in $\jump$. We now prove that $X$ is as
wanted by a series of claims.

\begin{itemize}
\item[(i)] $X$ is 1-generic. Indeed, at stage~$n$ of the construction,
either $\zeta_n$ is in $F_n$ (in fact in $F^{\min}_n$) or no extension
of $\zeta_n$ is in $F_n$. \\
\item[(ii)] In both cases (a) and (b) of the construction, we ensure
that no strict extension of $\zeta_n$ is in $F^{\min}_n$. Indeed
either case (a) holds, and $\zeta_n$ is itself in $F^{\min}_n$ in
which case no strict extension of $\zeta_n$ is, or case (b) holds, in
which case no extension of $\zeta_n$ is in $F_n$, and fortiori no
extension is in $F^{\min}_n$. Additionally, for all~$n$, $\zeta_n$ is
a strict extension of all $\zeta_k$ for $k<n$, therefore we conclude
by induction that for all~$n$ and all $k \leq n$, no strict extension
of $\zeta_n$ is in $F_k^{\min}$.\\
\item[(iii)] For all~$n$, $\permit(\zeta_n) \geq 4n+4$, and moreover,
any string $\sigma$ extending $\zeta_n$ by at least~$4$ bits satisfies
$\permit(\sigma) \geq 4n+8$. This is shown by induction. First, this
holds for $n=0$: all values of the function $\permit$ are greater or
equal to~$4$, in particular, $\permit(\zeta_0) \geq 4$. Now suppose
that $\permit(\zeta_n) \geq 4n+4$ for some~$n$. As we have seen in
claim (ii) above, no strict extension of $\zeta_n$ is in $F_k$ for any
$k \leq n$. Thus, for any extension $\zeta'$ of $\zeta_n$ and $a \in
\{0,1\}$ we have by definition of $\permit$: $\permit(\zeta' a) \geq
\min \{\permit(\zeta')+1, 4n+8\}$. From this, we see by a straightforward
induction that string $\sigma$ extending $\zeta_n$ by 4 bits or more
satisfies $\permit(\sigma) \geq 4n+8$. In particular, $\zeta_{n+1}$
extends $\zeta_n$ by at least~8 bits, hence $\permit(\zeta_{n+1}) \geq
4n+8$, which concludes the induction.\\
\item[(iv)] Similarly, for all~$n$, $\permit(\gamma_n) \geq 4n+4$.
This is true for $n=0$, and for $n>0$, since $\gamma_{n}$ is an
extension of $\zeta_{n-1}$ by 8 bits, it follows from (iii) that
$\permit(\gamma_n) \geq 4(n-1)+8 = 4n+4$.\\
\item[(v)] For all~$n$,  $M(\gamma_n) \geq \permit(\gamma_n)+8$. This
is true for $n=0$. For the induction step, we need to distinguish two
cases depending on how $\gamma_{n+1}$ was constructed from $\gamma_n$.
If we are in the above case (b), then $\gamma_{n+1}=\gamma_n
11111111$, and since $M(\gamma_n)>\permit(\gamma_n)$, $M$ bets and
wins 8 consecutive times, and thus $M(\gamma_{n+1})=M(\gamma_n)+16$.
Also, in that case $\permit(\gamma_{n+1}) \leq \permit(\gamma_n)+8$ by
definition of $\permit$ (adding one bit to a string can only increase
the value of $\permit$ by~$1$). From these two facts we can conclude
that $M(\gamma_{n+1}) \geq \permit(\gamma_{n+1})+8$. Suppose now that
$\gamma_{n+1}$ was constructed according to case~(a) above. In that
case, we need to precisely analyze the behavior of~$M$ and $\permit$
between $\zeta_n$ and $\gamma_n$, i.e. on strings of type $\zeta_n
\eta$ with $0 \leq |\eta| \leq |\gamma_n| - |\zeta_n|$. First,
$\zeta_n$ is an extension of $\gamma_{n-1}$ hence by (ii) no string
$\zeta_n \eta$ is in $F_k^{\min}$ for $k<n$. Additionally, since
$\gamma_n$ belongs to $F^{\min}_n$, no prefix of $\gamma_n$ does. This
shows by definition of $\permit$ that for $|\eta| \leq 4$ one has
$\permit(\zeta_n \eta)=\permit(\zeta_n)+|\eta|$ and for
$|\eta|\geq 4$ one has $\permit(\zeta_n \eta) \geq 4n+8$. On the
other hand, $M(\gamma_n) \geq \permit(\gamma_n)+8 \geq 4n+12$ (by
(iv)). Since $M$ can only decrease by~$2$ at each move, we have
$M(\zeta_n \eta) \geq 4n+6$ for any $|\eta| \leq 3$. But if
$|\eta|\geq 4$, as we just saw, the value of $\permit(\zeta_n
\eta)$ is at least $4n+8$, and the martingale~$M$ is never allowed to
bet if its capital is below $\permit$. Hence, it follows that
$M(\zeta_n \eta) \geq 4n+6$ whenever $0 \leq |\eta| \leq |\gamma_n| -
|\zeta_n|$. In particular, $M(\zeta_n) \geq 4n+6$, and since we are
in case (a), $\permit(\zeta_n) \leq 4n+4$, thus $M$ is allowed to bet and
wins 8 times consecutively, and $M(\gamma_{n+1}) \geq 4n+22$. Finally,
we have $\permit(\gamma_{n+1}) \leq \permit(\zeta_n)+8 \leq 4n+12$.
This finishes the induction.
\end{itemize}
We have seen in (i) that $X$ is 1-generic, and from (iv) and (v), it
follows that $\limsup_n M(\gamma_n) = +\infty$, hence~$M$ succeeds on~$X$. \qed
\end{proof}
The sequence~$X$ constructed in this last proof is 1-generic, hence by
Kurtz's result mentioned above~$X$ is also Kurtz random. We therefore
get the immediate corollary.

\begin{cor}\label{cor:kurtz-not-dm}
There exists a sequence~$X \in \cs$ which is Kurtz random but not
integer-valued random.
\end{cor}
The converse of this result is also true, that is there exists a
sequence~$X$ which is integer-valued random but not Kurtz random. To
prove this, we will need a different approach, via measure-theoretic
arguments, which we will present in Section~\ref{sec:bernoulli}.\\à

Strictly speaking, integer-valued martingales not only impose a lower limit on betting amounts but also require that all wagers be a multiple of the minimum bet.  We are therefore left with a question regarding the robustness of integer-valued randomness: if we remove the requirement that wagers must be a multiple of the minimum bet, do we still obtain the same notion of randomness?
\begin{open} \label{open:do we have to bet multiples}
Let $V$ be the set of all computable reals greater than or equal to 1 unioned with $\{0\}$.  Is $V$-valued random the same as integer-valued random?
\end{open}

\section{Finitely-valued martingales} \label{sec:fvm}

\noindent
We now consider the effects of imposing betting limits on martingale
strategies.  First we separate integer-valued randomness from
finitely-valued randomness.

\begin{thm}\label{thm:fin-random-not-int-random}
There exists an integer-valued martingale which succeeds on a
finitely-valued random.
\end{thm}

\begin{proof}
Partition the natural numbers into finite intervals, with $2^n$
intervals of length $n$ followed by $2^{n+1}$ intervals of length
$n+1$ for every $n$.  In a picture:
\[
I_{1,1}I_{1,2}I_{2,1} \dotsc I_{2,2^2} I_{3,1} \dotsc I_{3,2^3}
I_{4,1} \dotsc I_{4,2^4} I_{5,1}\dotsc
\]
where each interval $I_{n,\cdot}$ has length $n$.  Consider the class
of all sets $\mathcal{A}$ which guarantees that at least one ``1''
lies in each of these intervals.  An integer-valued martingale can
succeeds on any set in this class by using the ``classic'' martingale
strategy: in each interval bet \$1 on outcome ``1'', then bet \$2 on
outcome ``1'', then bet \$4 on outcome ``1'', etc.\ until the bet is
successful and then stop betting until the next interval.  In this
way, the gambler nets \$1 income over each interval.  After doing this
for each of the $2^n$ intervals of length $n$, she has enough money to
continue this strategy on the next intervals of length $n+1$.
Therefore some integer-valued martingale succeeds on every member of
$\mathcal{A}$.

On the other hand, we now find a $B \in \mathcal{A}$ on which no
finitely-valued martingale succeeds.  Let $M_0, M_1, M_2, \dotsc$ be a
list of all finitely-valued martingales.  Let $B(0) = 1$.  For
induction assume $B$ has been defined up through $I_n$, and try to
define $B$ on $I_{n+1}$ so that
\begin{itemize}
\item for some $e$, $M_e$ loses some money over $I_{n+1}$, and
\item for every $j<e$, $M_j$ gains no money over $I_{n+1}$.
\end{itemize}
If all the intervals $B$ are chosen so as to satisfy these
requirements, then all finitely-valued martingales will be
obliterated.  Indeed each index can only be chosen finitely many times
to play the role of $e$ before all the capital of $M_e$ is destroyed,
and therefore the choice of $e$ must go to infinity.

While it is impossible to choose values for $B$ so that these
requirements are satisfied on every interval, we can satisfy them
often enough to defeat every finitely-valued martingale.  Assuming
that $I_{n-1}$ has been built, we describe how to build $I_n$.  Recall
that a finitely-valued martingale always wagers integer dollar
amounts.  For each finitely-valued martingale $M$, let $\max(M)$
denote the maximum possible bet for $M$, and let
\[
L(e) = \sum_{j \leq e} \left[ \max(M_j)+1 \right].
\]
\begin{description}
\item{\emph{Claim:}}
Values for $B$ can be chosen in $I_n$ so that $M_e$ loses money if she
makes a nonzero wager before the last $L(e)$ positions of the interval
and is the lowest-indexed martingale to do so.  Furthermore for all
$j<e$, $M_j$ does not gain any money over $I_n$ with these values for $B$.

\end{description}
Thus if $M_0$ bets before the last $L(0)$ positions of $I_n$, $B$ can
force $M_0$ to lose money, thereby satisfying the construction
requirements.  So we need only consider the case where $M_0$ bets no
money before the last $L(0)$ positions of $I_n$.  By applying the
claim above inductively, we may assume that
\begin{itemize}
\item  each successive $M_e$ bets no money prior to the last $L(e)$
positions of $I_n$, and

\item for each $j \leq e$, $M_j$ earns no profit over $I_n$.
\end{itemize}
Eventually $B$ must encounter some martingale $M_s$ which is stupid
enough to bet money at the beginning of the interval, at which point
the requirements for $I_n$ can be satisfied (assuming $I_n$ is
sufficiently long to have such a ``beginning.'')   If $I_n$ is not
longer than $L(s)$ then the requirements are not satisfied on $I_n$.
But we do not worry about this failure because for all $e$ such that
$L(e) < \size{I_n}$, the way of choosing intervals prevents $M_e$ from
ever earning money again on any interval $I_{n+k}$ ($k \geq 0$).  Thus
for every $e$, there is a sufficiently large $N$ so that for all $n >
N$, $e$ gains no money from betting on $I_n$.  So $B$ defeats all
finitely-valued martingales.

It remains to prove the claim.  We argue by induction. Suppose that
$M_0$ makes a nonzero wager prior to the last $L(0)$ positions of the
interval $I_n$, say at position $x_0$.  We show how $B$ can force
$M(0)$ to lose money over $I_n$.  $B$ can act adversarially throughout
the interval except for the constraint inherited from the class
$\mathcal{A}$.  It follows that $M_0$'s betting amounts must be
nondecreasing from position $x_0$ until the end of the interval.  If
not, then $B$ can spend its obligatory ``1'' at the position where
$M_0$ decreased her bets.  $M_0$ already has a net loss at this point
of decrease, and $B$ can continue to act adversarially until the end.
 Therefore a decrease in betting amounts after $x_0$ would cause $M_0$
to lose.  Hence $M_0$ is forced to bet at least \$1 each of $L(0)$
times.  By the final bet in $I_n$, $M_0$ is already behind by at least
$\max(M_0)+1$, so this bet is irrelevant; $M_0$ has already lost.

Since we have already proved the claim when $M_0$ bets before the last
$L(0)$ positions, we can now focus on the case where $M_0$ bets only
during the last $L(0)$ positions.  Now it is easy to prevent $M_0$
from winning any money: $B$ places a ``1'' anywhere before the
$\size{I_n} - L(0)$ position and then $B$ can act adversarially on the
last $L(0)$ positions.  Any nonzero wager from $M_0$ will now
instantly result in a loss for $M_0$ because $B$ is free to everywhere
disagree with $M_0$.  Hence it suffices to consider the case where
$M_0$ does not bet anywhere and $B$ is obligated to post a ``1''
somewhere before the last $\size{I_n}-L(0)$ positions.

Curiously, $M_1$ now finds herself in exactly the same situation that
$M_0$ started with.  By same argument as above, $B$ can force $M_1$ to
lose money if $M_1$ bets prior to $\size{I_n}-L(1)$.  Therefore we can
reduce to the case where $M_1$ never bets and $B$ is obligated to
provide a ``1'' somewhere before $L(2)$.  The same argument holds for
$M_2, M_3, \dotsc$. Eventually some martingale $M_e$ has money and is
stupid enough to bet before $L(e)$.  At this point, the claim is proved.  \qed
\end{proof}

\begin{rem}
$B$ is $\jump$-recursive in the above construction.
\end{rem}

\noindent
Schnorr showed that for any set $A$, a real-valued martingale succeeds
on $A$ if and only if a rational-valued martingale succeeds on $A$ (see
\cite{Schnorr1971}, or \cite{Nies2009}~p.270).  His proof, however,
does not carry over to the finitely-valued case.

\begin{open}
If we allow finitely-valued martingales to bet real values instead of
rationals, do we get the same class of finitely-valued randoms?
\end{open}

\subsection{On single-valued randoms}

For the following discussion, it is useful to keep in mind that a real
is single-valued random if and only if it is $\{1\}$-valued random;
the particular dollar amount which is bet each round is immaterial.
For comparison with Kurtz randomness, we appeal directly to a theorem
of Doob (\cite{Doob1953}~p.324).  The following version for
``non-negative'' martingales appears in Ross's book (\cite{Ross1996},~p.316).

\begin{thm} \label{thm:doob convergence 1}
For every martingale $M$, the set of reals on which $M$ succeeds has
measure zero.  Furthermore, the capital of $M$ converges to some
finite value with probability 1.
\end{thm} 
Later, in Lemma~\ref{lem:defeat-ivm}, we shall appeal to a more
general version of Theorem~\ref{thm:doob convergence 1} (see
Billingsley \cite{Billingsley1995}~p.468).  A \emphdef{supermartingale process} a sequence $X_0, X_1, \ldots$ of real-valued random variables (possibly taking negative values) such that for all~$n$
\[
\mathbb{E}[X_{n+1} | X_0, X_1, \ldots, X_n]  \leq X_n.
\]
 \begin{thm}[Doob's Martingale Convergence Theorem] \label{thm:doob convergence 2}
Let $X_0, X_1, \dotsc$ be a supermartingale process (where each $X_i$
is a random variable).  If for some $m \in \mathbb{R}$, we have
$X_n \geq m$ for all $n$, then almost surely
$\lim_{n \to \infty} X_n$ exists and is finite.
\end{thm}


\begin{prop} \label{prop: Kurtz->svm}
Every Kurtz random is single-valued random.
\end{prop}

\begin{proof}
Suppose that some single-valued martingale $M$ succeeds on a real
$X$.  Let $\mathcal{F}$ denote the set of reals on which $M$ converges to
some finite value.  Then $X$ does not belong to $\mathcal{F}$, and $\mathcal{F}$ has
measure one by Theorem~\ref{thm:doob convergence 1}.  Hence $X$
belongs to the measure zero set $\complement{\mathcal{F}}$.

By definition of single-valued, $M$ is required to bet at every
position of the input real.  Hence the only way for $M$ to converge
to a finite value is to reach the value~$0$ and become constant. Therefore $\mathcal{F}$
is, in fact, the set of reals on which $M$ eventually goes broke. Thus
$\complement{\mathcal{F}}$ is the set of all infinite paths through the tree
$\{\sigma : M(\sigma)>0\}$.
It follows that $\complement{\mathcal{F}}$ is a recursive $\Pi^0_1$ class.  In
summary, $X$ belongs to a recursive $\Pi^0_1$ class of measure
zero and therefore is not Kurtz random. \qed
\end{proof}

\begin{rem}
The above argument shows even more: every Kurtz random is $V$-valued
random for any positive set of integers $V$.
\end{rem}

\noindent
As promised in the introduction, we now separate the incomparable
notions of bi-immunity, single-valued random, and law of large
numbers.  The reals satisfying the law of large numbers and the
bi-immune reals are already known to be different.  Their separation
can be deduced from the argument given in Proposition~\ref{prop:svm
not lln} with ``bi-immune'' substituted for ``single-valued random.''

\begin{thm} \label{thm: non-immune svm-random}
There exists a single-valued random which is neither immune nor co-immune.
\end{thm}

\begin{proof}
Let $M_0, M_1, \dotsc$ be a list of all possible $\{1\}$-computable
``gamblers,'' namely a list of pairs consisting of martingales and
their respective initial capital.  Define a set $A$ such that
\begin{itemize}
\item $A(n) = 1$ if $n \equiv 0 \mod 6$, and
\item $A(n) = 0$ if $n \equiv 3 \mod 6$.
\end{itemize}
The remaining values of $A$ work adversarially against the $M_i$'s.
Since the gamblers must bet exactly \$1 on each value of $A$, the
remaining values of $A$ can be chosen so as to force any particular
gambler to decrease his capital by a dollar over the course of any
three consecutive rounds of play.  We define the first initial segment
of $A$ so as to exhaust the capital of $M_0$, the following interval
of $A$ so as to exhaust the capital of $M_1$, etc.  Since each gambler
has only finite capital at any moment, each gambler's capital is
exhausted after a finite period of time.  Therefore no $\{1\}$-valued
martingale succeeds on $A$.  Furthermore, by the values assigned at
multiples of 3, $A$ contains an infinite recursive set as does its complement. \qed
\end{proof}

\begin{thm} \label{thm:bi-hyperimmune but not svm-random}
There exists a $\jump$-recursive, bi-hyperimmune set which is not
single-valued random.
\end{thm}

\begin{proof}
Let $\varphi_0, \varphi_1, \varphi_2, \dotsc$ be a list of the partial
recursive functions.  Define a $\jump$-recursive function $f$ satisfying
\[
(\forall e < n)\: [\varphi_e(f(n)) \converge \implies f(n+1) >
\varphi_e(f(n))+2]
\]
and let
\[
A = \{x : (\exists n)\: [f(2n) \leq x < f(2n+1)] \}.
\]
Now $A = \{a_0, a_1, a_2, \dotsc\}$ is bi-hyperimmune because for any
recursive $\varphi_k$,
\[
\varphi_k[f(2k+1)] < f(2k+2) - 2 \leq a_{f(2k+1)} - 1,
\]
and a similar inequality holds for the complement of $A$.  On the
other hand, the single-valued martingale strategy which bets on
$A(n+1)$ what the gambler saw at $A(n)$ will succeed on $A$.  This
strategy indeed succeeds because each time $A(n+1)$ disagrees with
$A(n)$, we have that $A(n+2)$ and $A(n+3)$ agree with $A(n+1)$.  So
over each three consecutive rounds of betting, the gambler increases
his capital by at least \$1. \qed
\end{proof}
\noindent
Proposition~\ref{prop: Kurtz->svm} and Theorem~\ref{thm:bi-hyperimmune but not svm-random} together give:
\begin{cor}
There exists a bi-immune set which is not Kurtz random.
\end{cor}

\begin{prop} \label{prop:svm not lln}
Single-valued random does not imply the law of large numbers and vice-versa.
\end{prop}

\begin{proof}
Unlike the set of reals which satisfy the law of large numbers, the
set of single-valued randoms is co-meager (by
Corollary~\ref{cor:comeagerness2}).  Moreover, the real
$.010101010\dotsc$ satisfies the law of large numbers but is not
single-valued random. \qed
\end{proof}
\noindent
Finally, we note that it is possible to separate single-valued
randomness from finite-valued randomness using an argument along the
lines of Proposition~\ref{thm: non-immune svm-random}.

\begin{prop} \label{prop:1,2-random}
There exists a $\{1,2\}$-valued martingale which succeeds on a
single-valued random.
\end{prop}

\begin{proof}
Similar to Theorem~\ref{thm: non-immune svm-random}, we partition the
natural numbers into intervals of length 5.  For the first two numbers
$n$ in each interval (that is, $n$ congruent to 0 or 1 (mod 5)), set
$A(n) = 0$ so that gambler can win at these places.  The last three
spots in each interval are adversarial against the single-valued
martingales.  $A$ will be able to defeat the $\{1\}$-valued martingale
since the best single-valued martingale strategy would first gain \$2
and then lose \$3 on each interval, for a net loss of \$1 per
interval.  Eventually the single-valued martingale will run out of
money.  On the other hand, there exists a $\{1,2\}$-valued martingale
which always bets \$2 on each of the first two numbers and \$1 on the
last three numbers in each interval, for a net gain of at least \$4 -
\$3 = \$1 per interval (regardless of any adversarial action that may
occur in the last 3 places).  Thus the money for this $\{1,2\}$-valued
martingale on $A$ goes to infinity. \qed
\end{proof}

\subsection{On \{0,1\}-valued randoms}

We can also separate single-valued randomness from finite-valued
randomness.
\begin{prop} \label{prop:{0,1}-bi-immune}
Let $V$ be any set containing $0$ and at least one other number~$n$.
Then any $V$-valued random is bi-immune.
\end{prop}

\begin{proof}
Let $A$ be a set which is not bi-immune; without loss of generality
assume that $A$ contains an infinite recursive set B.  Then a
$V$-valued martingale strategy which bets~$n$ dollars on members of
$B$ and $0$ on $A-B$ will succeed on $A$. \qed
\end{proof}

\noindent
The following corollary is a consequence of the definition of
finitely-valued random and Proposition~\ref{prop:{0,1}-bi-immune}.

\begin{cor} \label{cor:fvm->bi-immune}
finitely-valued random $\implies$ $\{0,1\}$-valued random $\implies$ bi-immune.
\end{cor}

\noindent
Since single-valued random does not imply bi-immune (Theorem~\ref{thm:
non-immune svm-random}), we obtain from Corollary~\ref{cor:fvm->bi-immune}:

\begin{cor}
There exists a $\{0,1\}$-valued random which is not single-valued random.
\end{cor}

\noindent
Although we were able to separate single-valued randomness from
$\{1,2\}$-valued randomness (Proposition~\ref{prop:1,2-random}), the
comparison between $\{0,1\}$-valued randoms and $\{0,1,2\}$-valued
randoms seems less clear.  We leave the reader with the following
interesting question.

\begin{open}
Is $\{0,1\}$-valued random the same as finitely-valued random?
\end{open}

\section{Integer-valued martingales and Bernoulli measures}
\label{sec:bernoulli}

In this last section, we present a proof of the fact that integer-valued randomness does not imply Kurtz randomness. We will get a counter example by choosing a sequence~$X$ at random with respect to some carefully-chosen probability measure.

Intuitively speaking, the Lebesgue measure~$\lambda$ on the
space~$\cs$ corresponds to the random trial where all bits are
obtained by independent tosses of a balanced 0/1-coin. An interesting
generalization of Lebesgue measure is the class of \emphdef{Bernoulli
measures}, where for a given parameter $\delta \in [-1/2,1/2]$ we
construct a sequence~$X$ by independent tosses of a coin with
bias~$\delta$ (that is, the coin gives~$1$ with probability
$1/2+\delta$ and $0$ with probability~$1/2-\delta$. This can be
further generalized by considering an infinite sequence of independent
coin tosses where the $n^\text{th}$ coin tossed has bias $\delta_n$.
This leads to the notion of \emphdef{generalized Bernoulli measures}. Formally,
on the space~$\cs$, given a sequence $(\delta_n)_{n \in \N}$ of
numbers in $[-1/2,1/2]$, a \emphdef{generalized Bernoulli
measure} of parameter~$(\delta_n)_{n \in \N}$ is the unique
measure~$\mu$ such that for all~$\sigma \in \fs$:
\[
\mu([\sigma]) = \prod_{n\; : \; \sigma(n)=0} (1-p_n)  \prod_{n\; : \;
\sigma(n)=1} p_n
\]
where $p_n = 1/2 + \delta_n$.  One can expect that if the $\delta_n$ are very small (that is,
$\delta_n$ tends to~$0$ quickly), then the generalized Bernoulli
measure of parameter~$(\delta_n)_{n \in \N}$ will not differ much from
Lebesgue measure. This was made precise by Kakutani.

\begin{thm}[Kakutani~\cite{Kakutani1948}]
Let $\mu$ be the generalized Bernoulli measure of
parameter~$(\delta_n)_{n \in \N}$. If the condition
\begin{equation}\label{eq:kakutani}
\sum_{n \in \N} \delta_n^2 < \infty
\end{equation}
holds, then $\mu$ is equivalent to Lebesgue measure $\lambda$, that
is, for any subset $\mathcal{X}$ of $\cs$, $\mu(\mathcal{X})=0$ if and
only if $\lambda(\mathcal{X})=0$. If condition~(\ref{eq:kakutani})
does not hold, then $\mu$ and $\lambda$ are inconsistent, that is,
there exists some~$\mathcal{Y}$ such that $\mu( \mathcal{Y})=0$ while
$\lambda(\mathcal{Y})=1$.
\end{thm}

If we want to work in a computability setting, we need to consider
\emphdef{computable} generalized Bernoulli measures, that is, those
for which the parameter $(\delta_n)_{n \in \N}$ is a recursive
sequence of reals. Vovk~\cite{Vovk1987} showed a constructive analogue
of Kakutani's theorem for computable generalized Bernoulli measures in
relation with Martin-L\"of randomness (perhaps the most famous
effective notion of randomness, but we do not need it in this paper).
The Kakutani-Vovk result has been used many times in the
literature~\cite{BienvenuM2009,MerkleMNRS2006,MuchnikSU1998,Shen1989}.
In particular, Bienvenu and Merkle proved the following.

\begin{thm}[Bienvenu and Merkle~\cite{BienvenuM2009}]
\label{thm:failure-kakutani2}
Let $\mu$ be a computable generalized Bernoulli measure of parameter
$(\delta_n)_{n \in \N}$. If $\sum_n \delta_n^2 = +\infty$, then the
class of Kurtz random sequences has $\mu$-measure~$0$.
\end{thm}

\noindent
To prove that integer-valued randomness does not imply Kurtz
randomness, we will construct a computable generalized Bernoulli measure~$\mu$
whose parameter $(\delta_n)_{n \in \N}$ converges to~$0$ sufficiently
slowly to have $\sum_n \delta_n^2 = +\infty$ (hence by the above
$\mu$-almost all sequences~$X$ are not Kurtz random, which we will
make even more precise) but sufficiently quickly to make~$\mu$ close
to Lebesgue measure and ensure that $\mu$-almost all sequences are
integer-valued random.

\begin{thm} \label{thm:dmr-not-kurtz}
There exists a sequence $X \in \cs$ which is integer-valued random but
not Kurtz random.
\end{thm}

\begin{proof}
We obtain~$X$ by choosing a random sequence with respect to the generalized Bernoulli measure of parameter $(\delta_n)$ with
\begin{equation*}\label{eq:parameter}
\delta_n = \frac{1}{\sqrt{n\; \ln n }}
\end{equation*}
for all $n>1$ (the values of $\delta_0$ and $\delta_1$ can be set
arbitrarily). We have $\sum_{i=2}^n \delta_i^2 \sim \ln \ln n$ (this
because $\int (t \ln t)^{-1} dt = \ln \ln t$, in particular $\sum_i
\delta_i^2 = +\infty$). By Theorem~\ref{thm:failure-kakutani2}, a
sequence~$X$ chosen at random according to the measure~$\mu$ will not
be (with probability~$1$) Kurtz random. We can even exhibit a
martingale~$M$ which wins against $\mu$-almost all sequences~$X$. It
is defined by $M(\emptystring)=1$ and for any string~$\sigma$ of length~$n$:
\[
M(\sigma 0)=(1-2 \delta_n)M(\sigma) \qquad \text{and} \qquad
M(\sigma 1)=(1+2 \delta_n)M(\sigma).
\]
This martingale is in fact the optimal martingale: when playing
against a sequence~$X$ that is chosen at random with respect to a
measure~$\nu$, the optimal martingale is defined by
$M(\sigma)=\nu([\sigma])/\lambda([\sigma])$. It is optimal in the
sense that for any other martingale~$M'$, we have for $\mu$-almost
all~$X \in \cs$: $M'(X \uh n)=O[M(X \uh n)]$ (see for
example~\cite{BienvenuM2009}). Here, if we take for $\nu$ our
generalized Bernoulli measure~$\mu$, the optimal martingale is exactly
the martingale~$M$. By Theorem~\ref{thm:failure-kakutani2}, for
$\mu$-almost all~$X$, $X$ is not Kurtz random, that is, there exists a
real-valued martingale $M'$ and a recursive order~$h$ such that
$M'(X \uh n) \geq h(n)$. But by
optimality, for any real-valued martingale $M'$ and $\mu$-almost
all~$X$, $M'(X \uh n)=O[M(X \uh n)]$. Putting all this together, there exists a recursive order~$h$ such that $M(X \uh n) \geq h(n)$ for all~$n$ and $\mu$-almost all~$X$.\\

However (and this will be crucial for the rest of the argument), $M$
succeeds quite slowly on average.  


\begin{lem}\label{lem:slow-win}
Let $r>0$ be a real number. Then for $\mu$-almost all~$X \in \cs$, one
has $M(X \uh n) = o(n^r)$.
\end{lem}

\noindent In order to prove this, we now see $X$ as a random variable
with distribution~$\mu$. We set for all~$n$:
\begin{equation} \label{eq:def-vn}
V_n = M(X \uh n)
\end{equation}
which is a martingale process. Then set
\begin{equation} \label{eq:def-ln}
L_n = \ln(V_n)
\end{equation}
By definition of~$M$ we have for all~$n$
\begin{equation}\label{eq:possible-vn}
V_{n+1} = \left\{ \begin{array}{l} (1+2\delta_n)V_n \qquad \text{with
probability $1/2+\delta_n$} \\ (1-2\delta_n)V_n \qquad \text{with
probability $1/2-\delta_n$} \end{array} \right.
\end{equation}
thus
\begin{equation}\label{eq:possible-ln}
L_{n+1} = \left\{ \begin{array}{l} L_n+ \ln(1+2\delta_n) \qquad
\text{with probability $1/2+\delta_n$} \\ L_n+\ln(1-2\delta_n) \qquad
\text{with probability $1/2-\delta_n$} \end{array} \right.
\end{equation}
Setting
\begin{equation} \label{eq:def-en}
e_n = \mathbb{E}[L_{n+1}-L_n] = (1/2+\delta_n)\ln(1+2\delta_n) +
(1/2-\delta_n)\ln(1-2\delta_n)
\end{equation}
(note \textit{en passant} that $e_n \sim 2\delta_n^2$ by same method
as \eqref{eqn:en passant}) we see that
\begin{equation} \label{eq:def-lnprime}
L'_n = L_n - \sum_{i=0}^{n-1} e_i
\end{equation}
is a martingale process. For all~$n$ we have $|L'_{n+1} - L'_n| \leq
e_n + 2\delta_n$ (here we use the fact that $\ln(1+x) \leq x$ for all
$x>-1$). We can thus apply Azuma's Inequality \cite{Azuma1967,Hoeffding1963,Ross1996} to $L'_n$: for all
integers~$n$ and positive real~$a$ one has
\begin{equation}\label{eq:azuma1}
\mu \{ L'_n \geq a \} \leq  \exp \left(  -\frac{a^2}{\sum_{i=0}^{n-1}
(e_i+2\delta_i)^2}  \right)
\end{equation}

Taking $a=r \ln n$ (for an arbitrarily small real $r>0$)
in~(\ref{eq:azuma1}) we get
\begin{equation}\label{eq:azuma2}
\mu \{ L'_n \geq r \ln n\} \leq \exp \left(  -\frac{r^2 (\ln
n)^2}{\sum_{i=0}^{n-1} (e_i+2\delta_i)^2}  \right)
\end{equation}

Since $e_i \sim 2\delta_i^2$, we have $e_i=o(\delta_i)$, so
\begin{equation}\label{eq:equiv1}
\sum_{i=0}^{n-1} (e_i+2\delta_i)^2 \sim \sum_{i=0}^{n-1} (2\delta_i)^2
\sim 2 \ln \ln n
\end{equation}

Thus for any~$n$ large enough:
\begin{equation}\label{eq:equiv2}
-\frac{r^2 (\ln n)^2}{\sum_{i=0}^{n-1} (e_i+2\delta_i)^2} \leq -2 \ln n
\end{equation}
Putting (\ref{eq:azuma2}) and (\ref{eq:equiv2}) together, we get
\begin{equation}\label{eq:azuma3}
\mu \{ L'_n \geq r \ln n \} \leq \frac{1}{n^2}
\end{equation}
for almost all~$n$. By the Borel-Cantelli lemma \cite{Ross1996}, since $\sum_n 1/n^2$
converges, with $\mu$-probability~$1$ the event $[L'_n \geq r \ln n]$
happens only finitely often, that is, with probability~$1$, for
any~$r>0$ and almost all~$n$, $L'_n \leq r \ln n$. Since
\[
L_n = L'_n + \sum_{i=0}^{n-1} e_i
\]
and $\sum_{i=0}^{n-1} e_i \sim 2 \ln \ln n$, it follows similarly
that, with $\mu$ probability~$1$, for any~${r>0}$ and almost all~$n$,
$L_n \leq r \ln n$. And as $L_n = \ln [M( X \uh n)]$, all this
entails that with $\mu$-probability~$1$, $M(X \uh n) \leq n^r$ for
any $r>0$ and almost all~$n$. This proves Lemma~\ref{lem:slow-win}.\\

Let $\ivm$ denote the class of integer-valued martingales.  We now
consider a restriction of integer-valued martingales: let $\ivm'$ be
the subset of $\ivm$, consisting of the integer-valued
martingales~$M$ that further satisfy $M(\sigma) < \sqrt{|\sigma|}$
for almost all~$\sigma$. The following lemma shows that the
martingales in $\ivm'$ are essentially as powerful as martingales in
$\ivm$ against sequences~$X$ chosen at random according to~$\mu$. \\



\begin{lem}\label{lem:ivm-ivmprime}
Let $M \in \ivm$. For $\mu$-almost all~$X$, there exists $M' \in
\ivm'$ such that $M'(X \uh n)  = M(X \uh n)$ for almost all~$n$.
\end{lem}

Let $M \in \ivm$. 
By Lemma~\ref{lem:slow-win}, $M( X \uh n) = o(\sqrt{n})$ almost surely. Hence, for
$\mu$-almost all~$X$, there exists some $n_0$ and all $n > n_0$, $M(X
\uh n) \leq \sqrt{n}/2$. For such a pair $(X,n_0)$, we call
``invalid'' all strings $\sigma$ such that there exists a prefix
$\tau$ of $\sigma$ such that $|\tau| \geq n_0$ and either $M(\tau 0)
> \sqrt{|\tau|}$ or $M(\tau 1) > \sqrt{|\tau|}$, and ``valid'' any
string that is not invalid. Now, define the martingale $M'$ by
$M'(\sigma)=M(X \uh n_0)$ for all $\sigma$ with $|\sigma| \leq n_0$
and for all $\sigma$ with $|\sigma| > n_0$, set $M'(\sigma)$ to be
$M(\tau)$ with $\tau$ the longest prefix of $\sigma$ that is valid.
In other words, $M'$ is the trimmed version of $M$ that stops
betting forever whenever $M$ makes at stage $n>n_0$ a bet that gives
it a chance to get a capital $> \sqrt{n}$. It is easy to see
that~$M'$ is itself a martingale, integer-valued as~$M$ is, and
since $M(X \uh n) \leq\sqrt{n}/2$ for all~$n > n_0$, all prefixes of $X$
are valid, hence $M'(X \uh n) = M(X \uh n)$ for all~$n \geq n_0$.
This proves the lemma.\\

Finally, we prove that martingales in $\ivm'$ are almost surely
defeated by a $\mu$-random~$X$.

\begin{lem}\label{lem:defeat-ivm}
Let $M \in \ivm'$. For $\mu$-almost all~$X$, $M$ does not succeed on~$X$.
\end{lem}

Let $n_0$ be such that $M(\sigma) \leq \sqrt{|\sigma|}$ for all
$\sigma$ of length $\geq n_0$. Again, we see~$X$ as a $\mu$-random
variable and define~$V_n$ by
\begin{equation} \label{eq:def-vn-bis}
V_n = M(X \uh n)
\end{equation}
(note that by definition of $\ivm'$, we have $V_n \leq \sqrt{n}$ for
all~$n\geq n_0$) and $L_n$ by
\begin{equation} \label{eq:def-ln-bis}
L_n =\ln[ M(X \uh n)]
\end{equation}
with the convention  $\ln(0)=-1$. For all~$n$, define also
\begin{equation} \label{eq:def-bn}
\rho_n = \frac{M(X \uh n+1) - M(X \uh n)}{M(X \uh n)}
\end{equation}
which is the fraction of its capital the martingale~$M$ bets on~$1$
at stage~$n$. It can be negative if~$M$ bets on~$0$ and is by
convention~$1$ if $M(X \uh n)=0$. Similarly
to~(\ref{eq:possible-ln}), we have for all~$n$:
\begin{equation} \label{eq:possible-ln-bis}
L_{n+1} = \left\{ \begin{array}{l} L_n+ \ln(1+\rho_n) \qquad
\text{with probability $1/2+\delta_n$} \\ L_n+ \ln(1-\rho_n) \qquad
\text{with probability $1/2-\delta_n$} \end{array} \right.
\end{equation}
Thus we have:
\begin{eqnarray}
\mathbb{E}[L_{n+1}-L_n] & = & (1/2+\delta_n)\ln(1+\rho_n) +
(1/2-\delta_n)\ln(1-\rho_n)\\
                                            & = & \frac{1}{2}
\ln(1-\rho_n^2) + \delta_n\ln(1+\rho_n) - \delta_n \ln(1-\rho_n)\\
                                            & \leq & -
\frac{\rho_n^2}{2} + 2 \delta_n \rho_n \label{eqn:en passant}
\end{eqnarray}
(for the last inequality, we use again that~$\ln(1+x) \leq x$ for
all~$x \geq -1$, which is true even for $x=-1$ with our convention $\ln(0)=-1$).
Now, observe that $\rho_n$ is either $0$, or of the form
$\frac{m}{V_n}$ for some integer~$m$ as~$M$ is integer-valued. In the
first case $L_{n+1}=L_n$ and in the second case, since $V_n \leq
\sqrt{n}$ for almost all~$n$, we have $|\rho_n|\geq 1/\sqrt{n}$
for almost all~$n$, and therefore $\mathbb{E}[L_{n+1}-L_n] \sim
-\frac{\rho_n^2}{2} < 0$ as $\delta_n=o(1/\sqrt{n})=o(\rho_n)$. This
shows that $L_n$ is ultimately a supermartingale, and it is bounded
from below by $\ln(0)=-1$. By Doob's Martingale Convergence Theorem
$L_n$ converges to a finite value $\mu$-almost surely, hence the same
is true for $V_n=\exp(L_n)$. Therefore~$V_n$ is $\mu$-almost surely
bounded, hence $M$ is $\mu$-almost surely defeated. This finishes the
proof of Lemma~\ref{lem:defeat-ivm}.\\

Theorem~\ref{thm:dmr-not-kurtz} now easily follows. Take some $X \in
\cs$ at random according to~$\mu$. By Lemma~\ref{lem:defeat-ivm}, $X$
defeats all~$M \in \ivm'$ $\mu$-almost surely, therefore by
Lemma~\ref{lem:ivm-ivmprime}, $X$ defeats all~$M \in \ivm$
$\mu$-almost surely. And finally, by definition of $\mu$ and
Theorem~\ref{thm:failure-kakutani2}, $X$ is $\mu$-almost surely not
Kurtz random. Therefore, $X$ is $\mu$-almost surely as wanted, hence
the existence of at least one~$X$ as wanted. \qed
\end{proof}

\section{Non-monotonic betting strategies}

A \emph{non-monotonic betting strategy} is a betting strategy in which
the gambler can bet on the bits of a sequence in any order she chooses
\cite{MerkleMNRS2006}.  A set $X$ is \emph{Kolmogorov-Loveland random}
if no recursive non-monotonic betting strategy succeeds on $X$ and \emph{Martin-L\"{o}f random} if no martingale with an increasing, recursive approximation succeeds on $X$.  By a theorem of Muchnik, Semenov, and Uspensky \cite{MuchnikSU1998} every Martin-L\"{o}f random is Kolmogorov-Loveland random, however the reverse containment remains a major open question for the field of
algorithmic randomness.

In the real-valued martingale case, every set which is computably
random relative to $K$ is also Martin-L\"{o}f random (unrelativized,
follows from martingale definition of ML-random) and hence
Kolmogorov-Loveland random. The situation is a bit different for the
case of integer-valued martingales.   We shall show that no oracle can
be given to a integer-valued martingale which will make it as powerful
as its non-monotonic counterpart.

\begin{thm}
For every oracle $B$ there is a set $A \leq_T B'$ and a
non-monotonic $\{0,1\}$-valued martingale such that the non-monotonic
martingale wins on $A$ while every $B$-recursive monotonic martingale
fails to win on $A$.
\end{thm}

\begin{proof}
Let $M_1,M_2,M_3,\ldots$ be a $B'$-recursive list of all
integer-valued $B$-recursive martingales with the additional
property that $M_m$ starts with at most $2^m$ dollars; note that
such a list can be made by following the $m^\text{th}$ program as long
as that program belongs to an integer-valued martingale with desired
properties and to freeze the martingale as constant (always betting
$0$) if at some time the $B$-recursive martingale turns out to be
partial or otherwise ill-defined. Note that although the $m^\text{th}$ program
might not be total, the $M_m$ are all total and uniformly $B$-recursive.

The idea is to construct a recursive partition $I_0,I_1,I_2,\ldots$
of intervals such that $I_n$ is so long that only the minority of the
positions in the interval are used to behave adversarially to
$M_1,M_2,\ldots,M_n$ while the majority of the $x$ in the interval
satisfies $A(x) = A[\min(I_n)]$. The basic idea is to select $A$
as follows on $x \in I_n$:
$$
  A(x) =
\begin{cases}
   b & \text{if $M_m$ bets a positive value on $1-b$ and} \cr
     & \text{all $M_k$ with $1 \leq k < m$ abstain from betting;} \cr
   0 & \text{if $x = \min(I_n)$ and} \cr
     & \text{$M_1,M_2,\ldots,M_n$ abstain from betting;} \cr
   A[\min(I_n)] & \text{if $x > \min(I_n)$ and} \cr
     & \text{$M_1,M_2,\ldots,M_n$ abstain from betting.} \cr
\end{cases}
$$
Note that the length of $I_n$ is determined although one does not know
anything about the martingales.
Let $a_1 = 2^{\min(I_n)+1}$ and inductively
$a_{m+1} = 2^{\min(I_n)+m+a_1+a_2+\ldots+a_m}$. The idea is that
$a_m$ stands for the largest value which $M_m$ can reach on the
interval $I_n$; the upper bound is determined by assuming that
$M_m$ --- in the worst case --- can double its capital whenever
the interval $I_n$ is not yet reached or that a martingale $M_k$
with $k < m$ is betting which is then given priority in the definition
of $A$. One can now verify by induction that each $M_m$ can bet on
$I_n$ only at most $a_m$ times until it would go broke and therefore
there are at most $a_1$ places where $M_1$ bets and loses $1$ out
of its capital; furthermore there are at most $a_2$ places where $M_2$
bets but $M_1$ does not bet and on these $M_2$ loses $1$ out of its
capital; there are at most $a_3$ places where $M_3$ bets inside $I_n$
and $M_1,M_2$ do not bet and on these places $M_3$ loses $1$ out of
its capital. In total there are at most $a_1+a_2+\ldots+a_n$ places
on $I_n$ where one of the martingales $M_1,M_2,\ldots,M_n$ are betting
and therefore by taking $I_n$ to have the length $2(1+a_1+a_2+\ldots+a_n)$
one gets that $A[\min(I_n)]$ coincides with the majority of the values
$A(x)$ with $x \in I_n-\{\min(I_n)\}$.

This property permits to implement a
non-monotonous recursive betting strategy which for every interval $I_n$
first reads all the values $A(x)$ with $x \in I_n-\{\min(I_n)\}$ without
betting any money on these values and then bets $1$ according to the
majority of the bits read before on the value $A[\min(I_n)]$;
this bet is correct and a sure win. Hence $A$ can be recognized by a
non-monotonous $\{0,1\}$-valued martingale.

Furthermore, one can find by induction values $x_1,x_2,\ldots$
such that $x_n \geq \min(I_n)$ and from $x_n$ onwards no martingale
$M_m$ with $m < n$ is betting on $A$. This is obviously possible for
$x_1 = \min(I_1)$ as the other part of the condition is void. Now,
whenever $M_1$ bets on $A$ beyond $x_1$, the outcome is negative
as $A$ gives highest priority to diagonalize $M_1$. Therefore,
$M_1$ can bet only finitely often until the capital is used off
and one can just take $x_2$ to be the maximum of $\min(I_2)$ and
the last time where $M_1$ places a positive bet on $A$. Hence $x_2$
exists and from $x_2$ onwards, $M_1$ does not bet on $A$ and therefore
$M_2$ is diagonalized with highest priority by $A$; again there are
only finitely many positive bets and $x_3$ can be chosen as the first
value after these finitely many bets and after $\min(I_3)$. Hence one
can inductively define the $x_n$ and verify that $M_n$ never has more
capital than $2^{x_n+1+n}$. Thus no $M_n$ succeeds on $A$ and $A$ is
integer-valued random (with respect to $B$-recursive monotonous
martingales). \qed
\end{proof}

We conclude with a canonical problem.

\begin{open}
Do there exist other characterizations for integer-valued,
finite-valued, or single-valued randoms in terms of Kolmogorov
complexity or Martin-L\"{o}f statistical tests?
\end{open}

\bibliographystyle{plain}
\bibliography{integer-valued-martingales}

\end{document}